# Retrocausal Quantum Effects from Broken Time Reversal Symmetry

Rajat Kumar Pradhan
Bhadrak Autonomous College, Bhadrak,
Odisha, India-756100

**Abstract**

Quantum effects arising from manifestly broken time-reversal symmetry are investigated using time-dependent perturbation theory in a simple model. The forward time and the backward time Hamiltonians are taken to be different and hence the forward and backward amplitudes become unsymmetrical and are not complex conjugates of each other. The effects vanish when the symmetry breaking term is absent and ordinary quantum mechanical results such as Fermi Golden rule are recovered.
PACS: 03.65Ta,

## 1. Introduction

Time reversal invariance has been a contentious issue [1,2] in non-relativistic quantum mechanics since its first description given by Wigner[3]. The Schrodinger equation $i\hbar(\partial \psi / \partial t) = H\psi$ is not invariant under $t \to -t$ and for conservation of transition probabilities requires it to be taken along with complex conjugation. Due to the hermiticity of the Hamiltonian the conjugate Schrodinger equation $-i\hbar(\partial \psi^* / \partial t) = H\psi^*$ represents the evolution of the conjugate state in backward time. But, in standard QM, both $\psi$ and $\psi^*$ are always treated on equal footing as they contain identical information about the system, though $\psi^*$ is hardly ever given an independent and explicit interpretation separately from $\psi$, except in Cramer's transactional interpretation [4]. Aharonov, Bergmann and Lebowitz [5] developed the time-symmetric version of QM, called the two state vector formalism (TSVF) using the forward evolving state $|\phi>$ and backward evolving quantum state $<\psi|$ as equal players in the determination of probabilities of measurement of an observable Q by the ABL rule:

$$\Pr(Q = q_n) = \frac{|<\psi | P_n | \phi >|^2}{\sum_j |<\psi | P_j | \phi >|^2} \quad \ldots (1)$$

This formula reduces to the usual Born rule of standard QM when there is no post-selection. Here the state of the system is described completely by the two-state vector $<\psi||\phi>$ and $P_j = |\phi_j><\phi_j|$ is the usual projection operator for the $j$th state[6].

A causally symmetric Bohm model has been proposed by Sutherland [7] wherein time-symmetry is utilized to explain quantum non-locality while maintaining Lorentz invariance. Time reversal symmetry however is contrary to our experience since we remember the fixed past and can only surmise on the uncertain future, and hence the forward-evolving physical state $|\phi>$ and the backward-evolving conjugate state $<\psi|$ cannot have equal significance. The entropic, cosmological and psychological arrows of time do point to manifestly broken time reversal invariance in nature and so do the CP-violating weak interactions, though the magnitude of the effect is very small in the latter case. Effects of PT symmetric non-hermitian interactions that violate P as well as T symmetry have also been studied in the literature [8] in various systems.

In this note, we study the effects of manifestly breaking time-reversal invariance using standard time-dependent perturbation theory by introducing a small T-breaking coefficient in the interaction term for the backward-evolving states. It turns out that retro-causation can be seen to be the effect (rather than the cause) of non-locality at a more fundamental level.

## 2. Breaking T-invariance by hand

Let the general physical state $|\phi(t_i,t)>$ for a system evolve forward in time from initial time $t_i$ by the forward-evolution Hamiltonian $H_F = H_0 + H'$ while the general backward evolving state $<\psi(t_f,t)|$ evolves by the backward evolution Hamiltonian $H_B = H_0 + (1+\lambda)H' = H_F + \lambda H'$ from a final time $t_f$ where, $\lambda$ is a small real-valued (in general time-dependent) dimensionless parameter that determines the extent of T-violation. Standard time-dependent perturbation theory of QM will be recovered when $\lambda = 0$. $H_0$ is the unperturbed Hamiltonian of the system having orthonormal eigen states defined by: $H_0 |n> = E_n |n>$. Note that both $H_F$ and $H_B$ are self-adjoint but they are not adjoints of each other, precisely because of the presence of the T-violating parameter $\lambda$ via the additional interaction term in $H_B$.

Such distinct evolutions by different forward and backward Hamiltonians have been studied by Hahne [9] using direct sum of the forward and backward Hilbert spaces as the state space. Here we examine the effects of introducing a time-dependent (in general) parameter $\lambda$ in the perturbation Hamiltonian for the backward evolution, somewhat as a simple hidden variable, which affects the quantum mechanical transition probabilities in a retrocausal manner.

Our aim is to find out the probability that if the system was in a given eigenstate $|i>$ of $H_0$ at $t_i$, what is the probability that it will be found in the eigenstate $|f>$ at time $t_f$ due to the different evolutions of the forward and backward evolving states. Further, using its dependence on the T-violating parameter $\lambda$, can we bring in a reasonable change in the spectrum of transition probabilities, thereby reducing quantum indeterminism? We consider some simple applications.

## 3. Modified Transition Probabilities

The transition probability in standard QM is calculated by the applying Born rule viz. taking modulus squared of the amplitude for the forward transition:

$$\Pr(i \to f) = \text{Amp}(i \to f) \times \{\text{Amp}(i \to f)\}^* \quad \ldots (2)$$

In view of T-symmetry in standard QM, we can write the backward transition amplitude as:

$$\text{Amp}(f \to i) = \{\text{Amp}(i \to f)\}^* \quad \ldots (3)$$

And, hence the probability can be written as:

$$\Pr(i \to f) = \text{Amp}(i \to f) \times \text{Amp}(f \to i) \quad \ldots (4)$$

In the model considered here, since the forward and backward amplitudes are not in general conjugates of each other due to broken T-symmetry, there will be a $\lambda$-dependence of the probabilities. Following Cramer[10], this can be explained as stemming from the interaction of the system with the backward travelling advanced waves (Confirmation echoes) from the future state, which can affect the transition probabilities during the interval $[t_i, t_f]$.

The forward amplitude for $i \neq f$ and to first order in the interaction $H'$, is given by[11]:

$$\text{Amp}(i \to f) = \frac{1}{i\hbar} \int_{t_i}^{t_f} e^{i\omega_{fi} t'} H'_{fi}(t') dt' \quad \ldots (5)$$

where, $\omega_{fi} = E_f - E_i$ and $H'_{fi}(t') = <f|H'(t')|i>$ is the matrix element of the interaction $H'$ connecting the initial and final states in the forward time direction.

Following the same way, the backward amplitude is given by:

$$\text{Amp}(f \to i) = -\frac{1}{i\hbar} \int_{t_f}^{t_i} e^{-i\omega_{fi} t'} \{1 + \lambda(t')\} H'^*_{if}(t') dt' \quad \ldots (6)$$

Using eq. (4), the probability then becomes:

$$\Pr(i \leftrightarrow f) = \Pr_{QM} + \Pr_{retro}(\lambda) \quad \ldots (7)$$

where the first term is the standard quantum mechanical probability for the transition while the second term is the additional retrocausal $\lambda$-dependent contribution to the probability. For this reason, the argument on the LHS has been signified with a left-right arrow. Some special cases of interest can now be considered:

(a) If $\lambda$ is a constant independent of time, then the probability becomes:

$$Pr(i \leftrightarrow f) = (1+\lambda) Pr_{QM} \quad \ldots(8)$$

If we can somehow have control over the parameter $\lambda$, we can deselect final states $|f'>$ other than the single final state $|f>$ by choosing $1+\lambda_{f'} = 0$ for all such states, thereby maximizing the probability of, and selecting, the state $|f>$ by retrocausal means.

(b) If $H'$ is a constant perturbation turned on at $t_i = 0$, then the probability is:

$$Pr(i \leftrightarrow f) = Pr_{QM} + \frac{|H'_{fi}|^2}{i\hbar(E_f - E_i)} (1 - e^{i\omega_{fi}t_f}) \int_{t_f}^{0} \lambda(t')e^{-i\omega_{fi}t'} dt' \quad \ldots(9)$$

Now, if $\lambda$ does not depend on time, then the formula again reduces to (8) with $Pr_{QM}$ given by the well-known oscillatory formula:

$$Pr_{QM} = \frac{4|H'_{fi}|^2}{|E_f - E_i|^2} \sin^2\left[\frac{(E_f - E_i)t_f}{2\hbar}\right] \quad \ldots(10)$$

From eq. (8) and eq. (10), one then obtains a modified Fermi golden rule containing the multiplicative factor $(1+\lambda)$, for the transition rate to the state $|f>$ within the group of states $\{f\}$ with energies nearly equal to the initial energy $E_i$ and having density of states $\rho(E_f)$:

$$w_{i\leftrightarrow f} = (1+\lambda) w_{i\rightarrow f} = (1+\lambda_f) \frac{2\pi}{\hbar} |H'_{fi}|^2 \rho(E_f) \quad \ldots(11)$$

where, in the last step we have introduced the state dependence of $\lambda$ by writing it as $\lambda_f$ to signify future state selection.

(c) For a harmonic perturbation of the form: $H' = Ve^{i\omega t} + h.c.$ turned on at $t_i = 0$ and with constant $\lambda$, the transition probabilities for emission $(E_f = E_i - \hbar\omega)$ and absorption $(E_f = E_i + \hbar\omega)$ are given respectively by:

$$w_{i\leftrightarrow f} = (1+\lambda) w_{i\rightarrow f} = (1+\lambda_f) \frac{2\pi}{\hbar} |V_{fi}|^2 \delta(E_f - E_i \pm \hbar\omega) \quad \ldots(12)$$

This formula is also applicable to find the transition probabilities for electric dipole transitions for an atom interacting with an applied electromagnetic field.

**4. Discussion**

In the above simple extension of quantum mechanical perturbation theory, we have interpreted the conjugate amplitudes as the backward time (retrocausal) amplitudes for a process by introducing a retro-causality parameter $\lambda$. We have shown that if the parameter is independent of time, then the transition probabilities are modified and the probabilities remain real and we have true retrocausal influences on the system. However, if $\lambda$ is time-dependent, then as is evident from eq. (9), the standard quantum mechanical formulae will be modified non-trivially depending on the exact nature of the dependence and probabilities will not remain real and will have an additional imaginary part which is difficult to interpret. It has been argued [Sutherland] that negative probabilities can be accommodated as long as the system is in transit, and when it approaches a measurement instant, the probabilities return to the interval [0,1]. This argument can be applied to cases in which some states are deselected by choosing $\lambda_f = -1$, so that the probability for the retro-causally selected state becomes ~1.

## 5. Conclusion

The validity of the model depends on whether we are able to detect retrocausal influences and whether the parameter $\lambda$ can be controlled by some means. For this, we must have temporal non-locality in some sense, since the final state must be known with greater degree of certainty in advance in order for us to influence the system in the backward time sense from the future. This in some sense has already been investigated [12] and encouraging results have been obtained using weak measurements [13] in the TSVF. In the model discussed here which is in terms of standard quantum mechanical perturbation theory, the uncertainty of the future state must correspondingly decrease as signified by the parameter $\lambda$ becoming ~1 for that state and ~0 for the rest of the states. There must be probability flows from rest of the final states to the intended one making it more certain as an outcome than when $\lambda$ is absent. It turns out that causal symmetry by itself cannot explain "true" retrocausal influences, which bring in more certainty of the realisation of the state. In contrast, the causal symmetry in the transactional model, Sutherland's bohemian model as well as in the TSVF will always keep intact the quantum mechanical probability assignments. Truly retrocausal influences via some kind of breaking of the T-symmetry as attempted here opens up new possibilities. How to exploit this is a matter to be taken up in future work.